\documentclass[]{aastex631}
\usepackage{url}
\usepackage{xcolor}
\usepackage{CJK}
\usepackage{amsmath,amssymb}
\usepackage{subfigure}

\shorttitle{Timing and Spectral Analysis of HMXB 4U 1700-37}
\shortauthors{Xiao et al.}

\graphicspath{{./}{figures/}}

\begin{document}

\title{Timing and Spectral Analysis of HMXB 4U 1700-37 Observed with \textit{Insight}-HXMT}

\author{Hua Xiao}
\affiliation{School of Physics and Astronomy, Sun Yat-sen University, Zhuhai, 519082, People's Republic of China}

\author[0000-0001-9599-7285]{Long Ji\textsuperscript{*}}
\email{jilong@mail.sysu.edu.cn}
\affiliation{School of Physics and Astronomy, Sun Yat-sen University, Zhuhai, 519082, People's Republic of China}

\author{Peng Zhang}
\affiliation{College of Science, China Three Gorges University, Yichang 443002, People's Republic of China}

\author{Lorenzo Ducci}
\affiliation{Institut für Astronomie und Astrophysik, Kepler Center for Astro and Particle Physics, Eberhard Karls, Universität, Sand 1, D-72076 Tübingen, Germany}
\affiliation{ISDC Data Center for Astrophysics, Universit\'e de Gen\`eve, 16 chemin d'\'Ecogia, 1290 Versoix, Switzerland}
\affiliation{INAF -- Osservatorio Astronomico di Brera, via Bianchi 46, 23807 Merate (LC), Italy}

\author{Victor Doroshenko}
\affiliation{Institut für Astronomie und Astrophysik, Kepler Center for Astro and Particle Physics, Eberhard Karls, Universität, Sand 1, D-72076 Tübingen, Germany}

\author{Andrea Santangelo}
\affiliation{Institut für Astronomie und Astrophysik, Kepler Center for Astro and Particle Physics, Eberhard Karls, Universität, Sand 1, D-72076 Tübingen, Germany}

\author{Shu Zhang}
\affiliation{Key Laboratory of Particle Astrophysics, Institute of High Energy Physics, Chinese Academy of Sciences, 19B Yuquan Road, Beijing 100049, People's Republic of China}

\author{Shuang-Nan Zhang} 
\affiliation{Key Laboratory of Particle Astrophysics, Institute of High Energy Physics, Chinese Academy of Sciences, 19B Yuquan Road, Beijing 100049, People's Republic of China}

\begin{abstract}
We report timing and spectral studies of the high-mass X-ray binary 4U 1700-37 using $Insight$-HXMT observations carried out in 2020 during its out-of-eclipse state. 
We found significant variations in flux on a time-scale of kilo-seconds, while the hardness (count rate ratio between 10-30\,keV and 2-10\,keV) remains relatively stable.
No evident pulsations were found over a frequency range of $10^{-3}$-2000\,Hz.
During the spectral analysis, for the first time we took the configuration of different $Insight$-HXMT detectors' orientations into account, which allows us obtaining reliable results even if a stable contamination exists in the field-of-view.
We found that the spectrum could be well described by some phenomenological models that commonly used in accreting pulsars (e.g., a power law with a high energy cutoff) in the energy range of 2-100\,keV.
We found hints of cyclotron absorption features around $\sim$ 16\,keV or/and $\sim$ 50\,keV.
\end{abstract}

\keywords{High mass x-ray binary stars (733), Accretion (14)}

\section{Introduction \label{sec:intro}}

High-mass X-ray binaries (HMXBs) are binary systems consisting of a compact object (a neutron star or a black hole) and a OB companion star. 
They can be further divided into Be/X-ray binaries and supergiant X-ray binaries (SgXBs). 
In SgXBs, the compact object accretes matter through strong winds and/or Roche-lobe overflow of stellar type O or B donors, and emits powerful X-rays \citep[for a review, see][]{Kretschmar2021}. 
The properties of the compact objects influence the accretion geometry. 
%
For example, if the accretor is a highly-magnetized neutron star (NS here and elsewhere), within the magnetosphere the accreting matter will move along the magnetic lines, and eventually falls onto the NS's surface near the polar caps, resulting in pulsations.
%
%

4U 1700-37 is an eclipsing high mass X-ray binary discovered by the $Uhuru$ satellite in December 1973 \citep{Jones(1973)}. 
It is known to be a prototypical wind-fed system, composed of an O6Iafpe supergiant donor star (HD 153919)  \citep{Hainich(2020)} and a compact object with an orbital period of 3.411581(7) days \citep{Falanga(2015)}
and a mass of 2.44 $\pm$ 0.27 $M_{\rm \odot}$ \citep{Clark2002}.
This mass is quite large for a neutron star, but relatively small to be a black hole. 
%
%
Although many observations have been made, the nature of 4U 1700-37 remains unclear. 
%
%
Pulsations were proposed with periods of 24\,min and 97\,min using $Copernicus$ and SAS-3 observations, respectively \citep{Branduardi(1978), Matilsky(1978)}.
However, they could not be confirmed by subsequent studies and \citet{Hammerschlag-Hensberge(1979)} pointed out that the 97\,min period could be a spurious signal due to instrumental artifacts. 
\citet{Murakami(1984)} detected a possible pulsation with a period of 67.4 s during the brightest flare using $Tenma$. 
After that, almost all studies on the pulsation searching in 4U 1700-37 failed, except for the discovery of mHz quasi-periodic oscillations (QPOs) \citep[e.g.][]{Martinez-Chicharro(2018),Jaisawal(2015),Boroson(2003)}. 
Based on the absence of pulsations and the discovery of the hard tail in the spectrum, \citet{Brown(1996)} suggested the nature of the compact object in 4U 1700-37 is a black hole.
However, we note that it is also possible that the pulsation is intermittent and only active within a short interval, like the case in LS +61$^{\circ}$303 \citep{Weng2022}.
In addition, \citet{Seifina(2016)} found that the spectrum of 4U 1700-37 can be described by two Comptonized components, and the dependence of the spectral index on the mass accretion rate and the seed photon temperature suggested the presence of a neutron star.

Another diagnostic for neutron stars is the detection of cyclotron resonant scattering features (CRSFs), which are shown as broad absorption lines in spectra.
If detected, they can be used to estimate magnetic fields of neutron stars as $E_{\rm cyc} \approx 11.6 \times B_{12} (1 + z)^{-1}$ (keV), where $E_{\rm cyc}$ is the centroid line energy, $B_{12}$ is the magnetic ﬁeld in the unit of 10$^{12}$ Gauss and $z$ is the gravitational redshift \citep[for a review, see][]{Staubert2019}.
However, the existence of a CRSF in 4U 1700-37 is still controversial.
\citet{Reynolds(1999)} reported the possible presence of a cyclotron absorption line at $\sim$37\,keV in the \textit{BeppoSAX} spectrum, which is then consistent with \textit{Suzaku} observations although it is quite model-dependent \citep{Jaisawal(2015)}.
On the other hand, the \textit{NuSTAR} observation disfavors this line, and instead suggests an absorption feature at $\sim$16 keV \citep{Bala(2020)}. 
Therefore, it remains an open question whether and where a cyclotron line exists.

In this paper, we present broadband \textit{Insight}-HXMT observations of 4U 1700-37.
This paper is structured as follows: In Section \ref{sec2}, we describe details of the data reduction and show timing and spectral analysis in Section \ref{sec3}. In Section ~\ref{sec4}, we discuss the implications of results and interpretations.

\section{Observation and Data Reduction \label{sec2}}
The \textit{Hard X-ray Modulation Telescope} \citep[\textit{Insight}-HXMT; ][]{Zhang(2020)} is China's first X-ray astronomical satellite, consisting of three main scientific payloads: the High Energy X-ray Telescope \citep[HE;][]{Liu(2020)}, the Medium Energy X-ray Telescope \citep[ME;][]{Cao(2020)} and the Low Energy X-ray Telescope \citep[LE;][]{Chen(2020)}. 
It has a wide energy coverage in 1-250\,keV, i.e., 20-250\,keV for HE, 5-30\,keV for ME and 1-15 keV for LE. 
Each payload is composed of detectors with different Field of Views (FOVs) and orientations.
In this paper, we considered all HE detectors and LE/ME detectors with small FOVs for the sake of background estimation.

The data reduction was performed by using {\sc hxmtdas v2.05} and the calibration database {\sc caldb v2.06}\footnote{\url{http://hxmtweb.ihep.ac.cn/software.jhtml}}.
The good time intervals (GTIs) were selected according to the recommended criteria of the official user's guide\footnote{\url{http://hxmtweb.ihep.ac.cn/SoftDoc/648.jhtml}}:
(1) the elevation angle $\ge$ 10$^{\circ}$; (2) the pointing offset angle $\le$ 0.04$^{\circ}$; (3) the value of the geomagnetic cutoff rigidity $\ge$ 8\,GeV; (4) stay away from South Atlantic Anomaly (SAA) passage for at least 300\,s. 

A bright source GX 349+2 is 1.47$^{\circ}$ away from 4U 1700-37, which is within {\it Insight}-HXMT's FOVs (Figure~\ref{SkyMap}).
As a result, all LE detectors and 2/3 ME detectors were significantly contaminated.
However, we note that both LE and ME have three different orientations (referred to as 0, 1 and 2, and 0, 1 and 2 as deﬁned in documentation of the tool { \tt hprintd\_detid}.), and the detectors with different orientations have different effective areas for off-axis sources.
This allowed us to reconstruct the temporal and spectral information of our target, alleviating the influence of the contamination source.
In practice, we extracted the lightcurves and spectra from detectors that have the same orientation, as well as their corresponding response files for both 4U 1700-37 and GX 349+2.
Their backgrounds were estimated by using the tools {\tt\string lebkgmap} and {\tt \string mebkgmap} \citep{Liao(2020a)}, which were then re-normalized according to the selected detector IDs.
The observed spectra are the sum of contributions from two sources and the instrumental background:
\begin{equation}
   S_{\rm i}=M_{\rm A}\times R_{\rm i, A} + M_{\rm B}\times R_{\rm i, B} + B_{\rm i}, i=0, 1, 2
\end{equation}
where $S_{\rm i}$ and $B_{\rm i}$ represent the observed spectrum and the background for the $i$ orientation, $M_{\rm A}$ and $M_{\rm B}$ represent spectral models for 4U 1700-37 and GX 349+2 respectively, and $R_{\rm i, A}$ and $R_{\rm i, B}$ are their response matrices.
The spectral analysis can be done by fitting all spectra jointly using the two models for two sources.
{Based on the \textit{Insight}-HXMT's in-flight calibration, energy ranges of LE, ME, and HE detectors were limited to 2-8\,keV, 8-20\,keV and 30-100\,keV during the spectral fittings \citep{Li(2020)}.
In addition, the 20-30\,keV energy band was ignored due to calibration issues caused by sliver ﬂuorescent lines and the high background. 
The spectral analysis was performed by using the X-ray spectral fitting package {\sc xspec} version 12.12.0 \citep{Arnaud(1996)}.
$\chi^2$-statistics was used during fittings and a systematic uncertainty of 1 $\%$ was added.
All uncertainties quoted in this paper correspond to a 90$\%$ confidence level, calculated by running Monte Carlo Markov Chains (MCMC) with a length of 10000 and a burn-in of 10000 using the Goodman-Weare algorithm.

Similarly, lightcurves from both 4U 1700-37 and GX 349+2 can be resolved.
For a given time interval, the total count rate ($R_{\rm i}$) from detectors with the $i$ orientation should be the sum of the instrumental background ($R_{\rm back}$) and contributions from the two sources:
\begin{equation}
   R_{\rm i} = R_{\rm bkg} + R_{\rm A} + f_{\rm i} \cdot R_{\rm B}, i=0, 1, 2  
\end{equation}
where the variables $R_{\rm A}$ and $R_{\rm B}$ represent on-axis count rates of 4U 1700-37 and GX 349+2, and $f_{\rm i}$ is the ratio of the off-axis effective area to the on-axis case for the contamination source.
Here $f_{\rm i}$ can be determined from response files, since for \textit{Insight}-HXMT it is only a function of coordinates. 
The values of $f_{\rm 0}$, $f_{\rm 1}$, and $f_{\rm 2}$ considered here are 0.485, 0.083 and 0.362 for LE and 0.115, 0.0 and 0.244 for ME, respectively.
Finally, $R_{\rm A}$ and $R_{\rm B}$ can be fitted using the least square method by comparing three $R_{\rm i}$ with real observations.

\begin{figure}
  \includegraphics[width=0.9\linewidth]{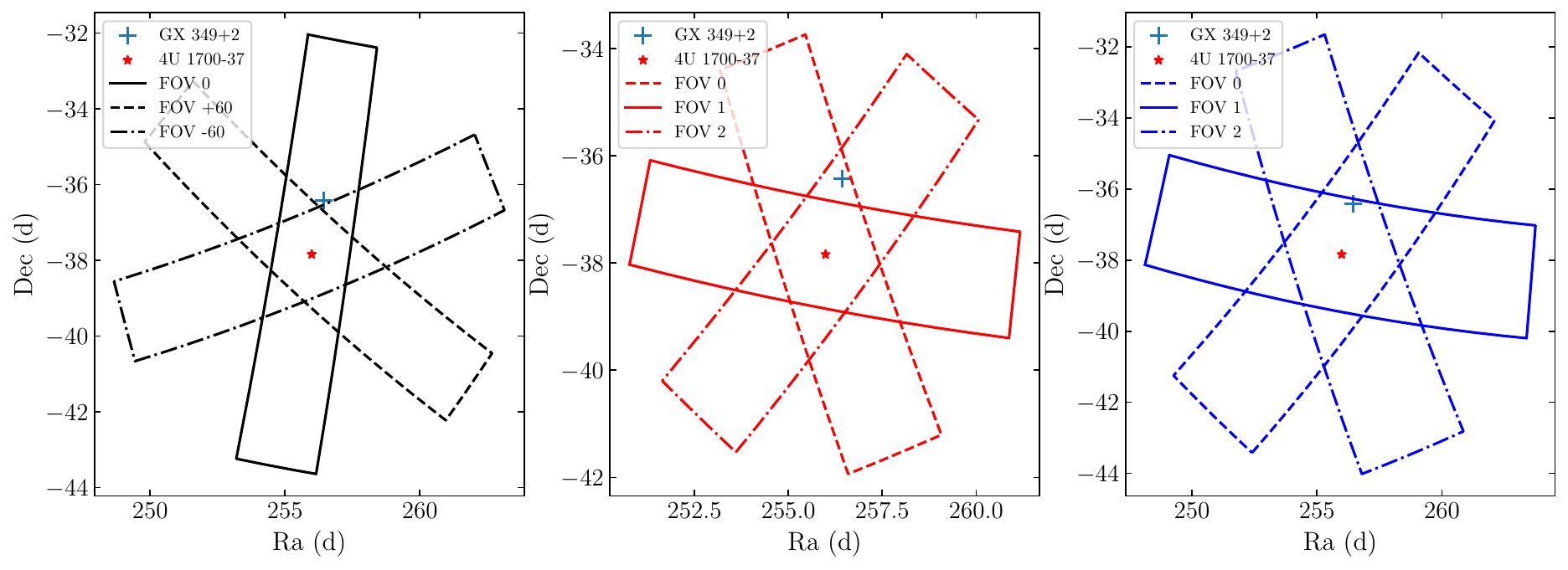}
  \centering
  \caption{HE (left), ME (middle) and LE (right) FOVs, where our target 4U 1700-37 is located in the center and the blue `+' symbol represents GX 349+2.\label{SkyMap}}
\end{figure}

\begin{figure}[ht!]
  \includegraphics[width=0.55\linewidth]{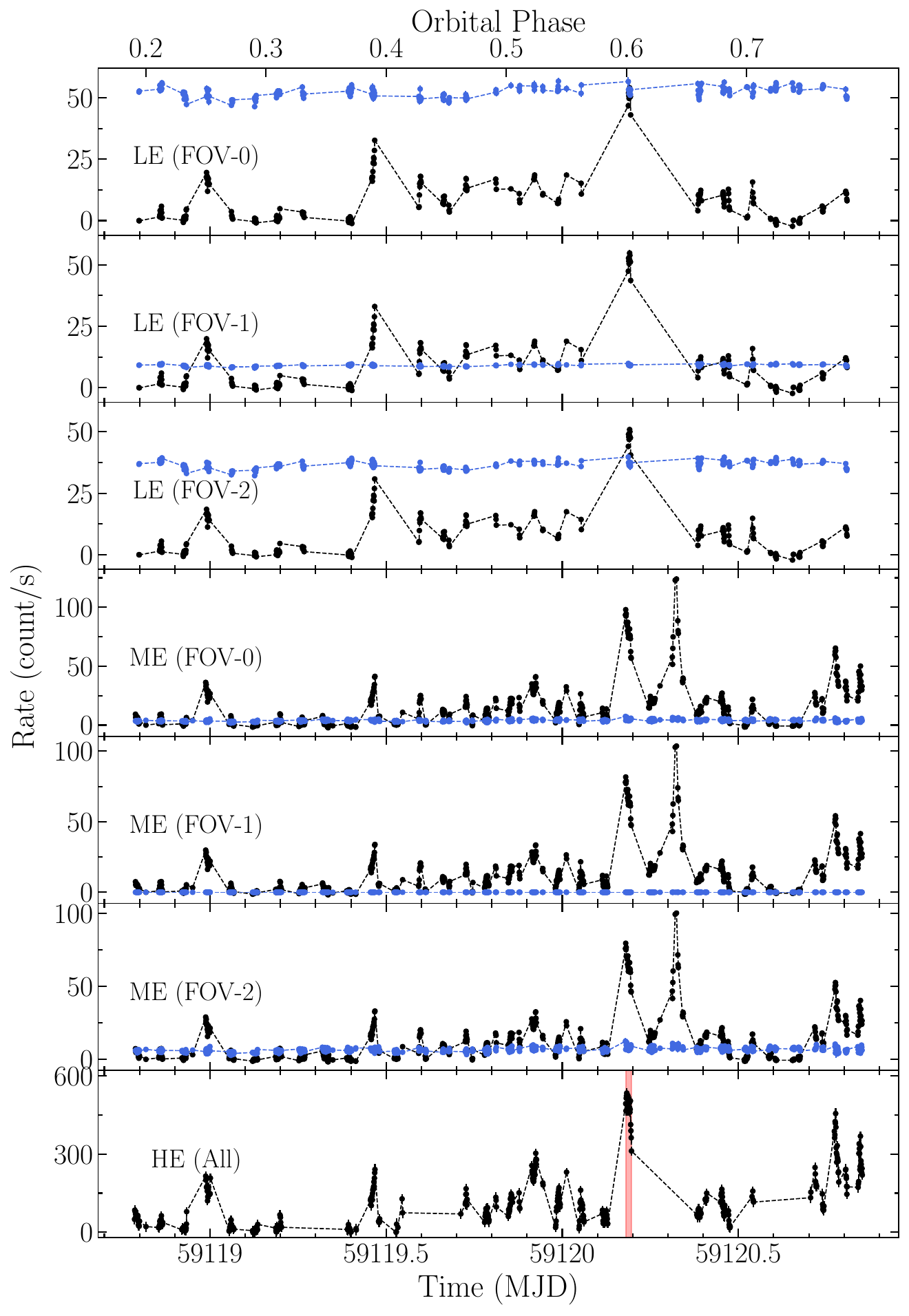}
  \centering
  \caption{
   The background-subtracted LE (Panel 1-3) and ME (Panel 4-6) lightcurves observed with $Insight$-HXMT with a time resolution of 64\,s, where black and blue lines present contributions from 4U 1700-37 and GX 349+2, respectively.
  Bottom panel: the background-subtracted HE lightcurve, which contains all available detectors, where the red region indicates the flare for which dedicated pulsation search was conducted in addition to search over the full dataset.
\label{lightcurve}}
\end{figure}

\section{Results} \label{sec3}

\subsection{Timing analysis}
4U 1700-37 was observed with \textit{Insight}-HXMT in September 2020 (MJD 59118.78-59120.85), spanning the out-of-eclipse orbital phase 0.19-0.78.
Here we adopted the ephemeris provided by \citet{Falanga(2015)}, where the orbital phase 0 corresponds to the mid-eclipse time.

Figure~\ref{lightcurve} shows backgrounds subtracted lightcurves in three different energy bands with a time resolution of 64\,s.
We considered all HE detectors since the influence of the contamination source was negligible because of its soft spectral shape \citep{Coughenour(2018)} 
.
On the other hand, LE lightcurves were calculated using the method mentioned in Section \ref{sec2}.
It is clear that 4U 1700-37 has a strong variability, especially exhibiting flares around the orbital phase 0.6, while the count rate of GX 349+2 is relatively stable during our observation.
As a comparison, Figure~\ref{lightcurve} also presents the ME lightcurve that only contains uncontaminated detectors, and indicates a very similar trend.
This, in turn, justifies the correctness of our lightcurve estimation method.

We show the hardness evolution in Figure~\ref{hardness}, where the hardness is defined as the count rate ratio between 10-30\,keV and 2-10\,keV.
In contrast to the strong variation in lightcurves, the hardness of 4U 1700-37 generally shows a marginal change throughout the observation, including the flaring episodes. 
A significantly increased hardness only appears near the ingress and egress phases, which is consistent with previous observations \citep[e.g.,][]{Jaisawal(2015)}.
In Figure~\ref{hardness}, we also present the hardness-intensity diagram (HID), which however does not show any obvious patterns as reported commonly in other X-ray binaries.

\begin{figure}[ht!]
	\includegraphics[width=1\linewidth]{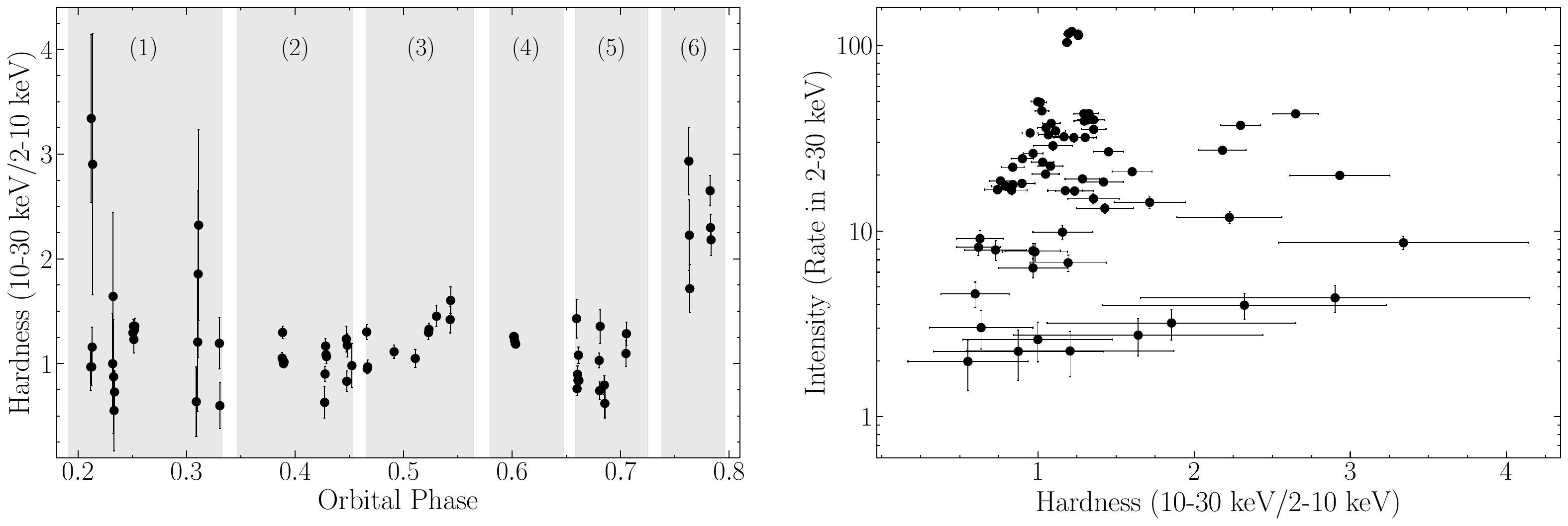}
	\centering
	\caption{Left panel: the hardness evolution of 4U 1700-37, where the hardness is defined as the count rate ratio between 10-30\,keV and 2-10\,keV. The grey stripes mark 6 epochs used in the following spectral analysis.
		%
		%
		%
		Right panel: the hardness-intensity diagram of 4U 1700-37.
		\label{hardness}}
\end{figure}


To search for periodic signals, we produced a power spectrum using HE lightcurves (Figure~\ref{powerspectrum}), calculated by the software {\sc stingray} \citep{Huppenkothen2019}.  
Here only HE was considered because of its high statistics and the negligible influence from the contamination source.
The power spectrum is dominated by a red noise at low frequencies, where an excess signal seems to appear around 7\,mHz.
However, we caution that its significance level is less than 99.9\%, estimated by Monte-Carlo simulations assuming that the red noise has a power-law shape \citet{Timmer1995}.
We also checked the lomb-scargle periodogram, which had similar results.
In addition, we inspected the power spectrum at higher frequencies (i.e., 1-2000\,Hz), and found that it was completely dominated by the white noise, with no periodic signals detected.
We also attempted to search for intermittent pulsations only during the flaring episode (the red region in Figure~\ref{lightcurve}), but no pulsations were found as well.

\begin{figure}
	\includegraphics[width=0.5\linewidth]{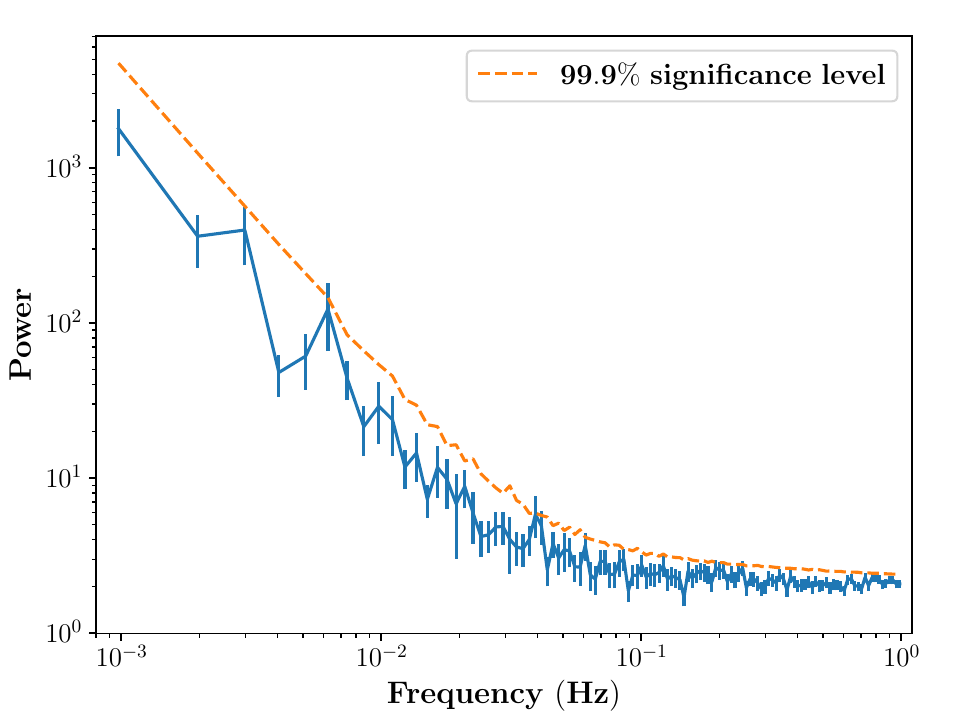}
	\centering
	\caption{The averaged power spectrum of \textit{Insight}-HXMT/HE's observation in the energy range of 30-100\,keV. The dashed line presents the 99.9\% significance level of the periodicity detection based on Monte-Carlo simulations.
 }
\label{powerspectrum}
\end{figure}

\subsection{Spectral analysis}
We performed the spectral analysis by jointly fitting spectra obtained from LE/ME detectors with different orientations (6 in total), and considered the contributions from both 4U 1700-37 and GX 349+2 simultaneously.
On the other hand, the contamination from GX 349+2 on high energies was negligible because of its soft spectral shape. Thus, HE spectra were only used to fit 4U 1700-37.

For 4U 1700-37, we first adopted two phenomenological models that are widely used in accreting pulsars \citep[for a review, see][]{Staubert2019}: {\tt\string cutoffpl} and {\tt\string highecut}. 
The former is a powerlaw model with a high energy exponential rolloff
$$I_{\rm E}=K \cdot E^{-\Gamma} {\rm exp}(-E/E_{\rm fold})$$
and the latter is a product of a powerlaw model and a multiplicative exponential factor
$$ I_{\rm E}=\left\{
\begin{aligned}
& K \cdot E^{-\Gamma},\ {\rm if}\ E \leq E_{\rm cut}\\
& K  \cdot E^{-\Gamma} {\rm exp}(-\frac{E - E_{\rm cut}}{E_{\rm fold}}),\ {\rm if}\ E \geq E_{\rm cut}\,.
\end{aligned}
\right.
$$
To account for the residuals at soft X-rays, an additional {\tt\string blackbody} component was also included.
According to previous studies \citep{Jaisawal(2015),Martinez-Chicharro(2018),Bala(2020),Martinez2021}, the spectrum of 4U 1700-37 is modified by the photoelectric absorption by partial covering materials.
Therefore, we employed the {\tt\string TBabs} model \citep{Wilms(2000)} to account for the Galactic absorption, where the equivalent hydrogen column ($N_{\rm H,1}$) was fixed at $\rm 0.5\times 10^{22} atoms\ cm^{-2}$ \citep{HI4PI2016}, and the {\tt\string TBpcf} model to describe the partial covering absorption of local materials.
A {\tt\string gauss} component was also added to fit the Fe fluorescence line, of which the width was fixed at 10\,eV because of the limited energy resolution of $Insight$-HXMT/LE.
In addition, to account for the cross calibration between different instruments, we used a {\tt\string constant} model that was fixed at 1 for ME/box-0 and free for other detectors.
For GX 349+2, we applied a spectral model of {\tt\string constant*tbabs(gauss + diskbb + bbodyrad)} as suggested by {\it NuSTAR} observations \citep{Coughenour(2018)}, where the {\tt\string diskbb} component presents a multi-color accretion disk \citep{Mitsuda1984}.
Here the {\tt\string constant} parameters were linked to 4U 1700-37's model that corresponds to the same spectra.

In the end, we found that both {\tt\string highecut} and {\tt\string cutoffpl}  models are able to describe the observed spectra successfully, which result in $\chi^2$/dof=2783.32/2762 and 2824.12/2763, respectively (Model 1 and Model 3 in Table \ref{model}).
We show the best-fitting parameters and residuals in Table~\ref{model} and Figure~\ref{Spectrum}.
For comparison, we tested a physical model {\tt\string nthcomp} \citep{Zdziarski1996,Zycki1999}, which is a thermally Comptonized model (see Model 5).
This also leads to an acceptable goodness-of-fit, with an asymptotic power-law photon index of $\Gamma=1.96$ and a electron temperature of $kT_{\rm e} \approx 16.33$\,keV.
If we assume a spherical geometry, the optical depth $\tau$ is estimated to be $\sim$ 3.7 by using the equation $\Gamma=-\frac{1}{2}+\sqrt{\frac{9}{4}+\frac{1}{\frac{kT_{\rm e}}{mc^2}\tau(1+\frac{\tau}{3})}}$ \citep{Zdziarski1996}.
Furthermore, to test the influence of GX 349+2 on the modeling of 4U 1700-37, we considered another spectral model commonly used in "Z"-sources, i.e., replacing the {\tt\string bbodyrad} component with a {\tt\string powerlaw} model.
In this case, we could also obtain good fitting qualities (Model 2 and 4 in Table~\ref{model}) with only a small variation in the best-fitting parameters of 4U 1700-37.
This means that, when taking the orientation configuration of \textit{Insight}-HXMT into account, the model selection of one source (if it works) has little influence on another one, reflecting the capability of the spectral analysis even if existing a contamination source in the FOV.

Although overall the goodness-of-fit (i.e., $\chi^2$) of these models is statistically acceptable, we note that there are still structures in the residuals around 50\,keV (especially for Model 5), which may hint at the presence of a cyclotron absorption feature. 
To test this possibility, we included an additional Gaussian absorption component ({\sc GABS} in {\sc XSPEC}), and estimated its significance via \textit{Akaike Information Criterion} \citep{Akaike1974} \mbox{${\rm AIC} = \chi^2 + 2k+(2k^{2}+2k)/(n-k-1)$},  where $k$ and $n$ are the number of free parameters of the model and the number of spectral bins, respectively. The probability of chance improvement was evaluated as $\rm exp(-\Delta_{AIC}/2)$ between the cases without and with the cyclotron absorption line.
We found that the significance level of this component is model-dependent, which is important for Models 3, 4 and 5 ($>5\sigma$\footnote{However, we note that only statistical errors were considered here. Around 50\,keV, the background starts to be important (see Figure \ref{background}), and its systematic error cannot be ignored.
So we should take these significance levels with caution.}).
The results are shown in Table~\ref{model2} and Figure~\ref{fig:residual}.

As a comparison, we also performed simulations to estimate the significance level, and obtained comparable results.
In practice, for each model we simulated $1.5\times10^4$ faked spectra assuming no additional component (i.e., the null hypothesis), and then fitted them using models with and without the component.
The significance level can be obtained by comparing the observed $\Delta\chi^2$ and the $\Delta\chi^2$ distribution from simulated samples.
%

We tested the CRSFs around 16\,keV and 37\,keV as reported by \citet[]{Reynolds(1999), Jaisawal(2015), Bala(2020)}.
Since our data cannot constrain both 50\,keV and 16 (or 37)\,keV lines together, we removed the 50\,keV component and considered the 16 (or 37)\,keV line only.
The width of the 16\,keV CRSF cannot be constrained, which was therefore fixed at 5\,keV according to \citet{Bala(2020)}.
The results are shown in Table~\ref{model3} and Figure~\ref{fig:residual}, which indicates that the detection of the 16\,keV cyclotron line is marginal and also model-dependent, which is well consistent with the result of \citet{Bala(2020)}.
Using the same strategy, we also tested the 37\,keV cyclotron line.
However, we did not find an evident detection ($>2\sigma$) even if we fixed the line width for all different models.
%
%



In order to investigate the evolution of 4U 1700-37 along with the orbital phase, 
we divided the entire observation into 6 epochs (see Figure \ref{hardness}) and performed a time-resolved spectral analysis.
The results are shown in Figure~\ref{Spectral_parameters}.
Here, for simplicity, we only chose the Model 1, i.e., using the {\tt\string highecut} model included a Gaussian absorption component at $\sim$ 16\,keV for 4U 1700-37 and the {\tt\string bbodyrad} model for GX 349+2.
We note that GX 349+2 was quite stable during the entire observation (see the lightcurve in Figure~\ref{lightcurve}). Therefore, its parameters were fixed at the time-averaged values.
In addition, the Fe(K$\alpha$) line and the Gaussian absorption parameters cannot be well constrained and also fixed at the averaged values due to the poor statistics.
As shown in Figure~\ref{Spectral_parameters}, the flux varies between 2.51 to 20.18 in units of $\rm10^{-9} erg\,s^{-1}\,cm^{-2}$, where the brightest epoch appears at the orbital phase 0.6 corresponding to the flaring state.
In contrast, the spectral shape generally does not present dramatic changes, consistent with the hardness evolution.
The only exception is the last epoch, where the absorption is increased significantly, accompanying with a decreased partial covering fraction and a decreased powerlaw index.

\begin{deluxetable}{ccccccc}
\tablecaption{Best-fitting parameters of 4U 1700-37 and GX 349+2. \label{para}}
\tabletypesize{\scriptsize}
\tablewidth{400pt}
\tablenum{1}
\renewcommand\arraystretch{1.4}
\tablehead{
\colhead{Source} & \colhead{Parameter} & \colhead{Model 1} & 
\colhead{Model 2} & \colhead{Model 3} & 
\colhead{Model 4} &
\colhead{Model 5} 
} 
\startdata
4U 1700-37&$\Gamma$&1.32$^{+0.04}_{-0.03}$&1.34$^{+0.04}_{-0.04}$&1.31$^{+0.03}_{-0.03}$&1.22$^{+0.03}_{-0.02}$& 1.96$^{+0.04}_{-0.03}$ \\
&${N_{\rm H,1}}$ (10$^{22}$ cm$^{-2}$)&0.50(fixed)&0.50(fixed)&0.50(fixed)&0.50(fixed)&0.50(fixed) \\
&$N_{\rm H,2}$ (10$^{22}$ cm$^{-2}$)&7.52$^{+0.83}_{-0.80}$&7.77$^{+0.24}_{-0.23}$&7.95$^{+1.15}_{-0.56}$&7.85$^{+0.20}_{-0.30}$& 3.08$^{+1.05}_{-0.74}$\\
&$f$&0.92$^{+0.03}_{-0.03}$&0.91$^{+0.01}_{-0.01}$&0.82$^{+0.01}_{-0.01}$&0.89$^{+0.02}_{-0.03}$&1.00$^{+0.00}_{-0.14}$ \\
&$E_{\rm line}$ (keV)&6.47$^{+0.02}_{-0.02}$&6.47$^{+0.02}_{-0.03}$&6.48$^{+0.02}_{-0.02}$&6.48$^{+0.02}_{-0.03}$&6.47$^{+0.03}_{-0.04}$ \\
&${\sigma_{\rm line}}$ (eV)&10.00(fixed)&10.00(fixed)&10.00(fixed)&10.00(fixed)&10.00(fixed) \\
&$kT_{\rm BB}$ (keV)&4.20$^{+0.15}_{-0.07}$&4.25$^{+0.10}_{-0.16}$&4.10$^{+0.07}_{-0.05}$&3.87$^{+0.08}_{-0.08}$&\nodata \\
&$Norm_{\rm BB}$ ($10^{-3}$)&305$^{+24}_{-45}$&294$^{+28}_{-36}$&454$^{+53}_{-52}$&460$^{+73}_{-35}$&\nodata \\
&$E_{\rm cut}$ (keV)&9.32$^{+0.63}_{-0.50}$&9.23$^{+0.53}_{-0.71}$&29.35$^{+1.10}_{-1.00}$&27.30$^{+0.88}_{-0.35}$&\nodata \\
&$E_{\rm fold}$ (keV)&29.01$^{+1.28}_{-1.02}$&29.53$^{+1.04}_{-1.02}$&\nodata&\nodata&\nodata \\
&$kT_{\rm e}$ (keV) &\nodata &\nodata &\nodata &\nodata &16.33$^{+1.29}_{-0.75}$ \\
&$kT_{\rm seed}$ (keV) &\nodata &\nodata &\nodata &\nodata &2.20$^{+0.06}_{-0.05}$ \\
&$Norm_{\rm comp}$ ($10^{-3}$) &\nodata &\nodata &\nodata &\nodata &8.26$^{+0.43}_{-0.30}$ \\
&$Flux$ (10$^{-9}$\,erg\,cm$^{-2}$ s$^{-1}$)&6.82$^{+0.11}_{-0.00}$&6.83$^{+0.07}_{-0.07}$&7.46$^{+0.07}_{-0.05}$&6.86$^{+0.09}_{-0.05}$&6.69$^{+0.07}_{-0.10}$ \\
&$C_{\rm LE0}^*$&0.80$^{+0.01}_{-0.01}$&0.81$^{+0.01}_{-0.02}$&0.89$^{+0.01}_{-0.02}$&0.80$^{+0.01}_{-0.01}$&0.76$^{+0.01}_{-0.02}$ \\
&$C_{\rm LE1}$&0.83$^{+0.01}_{-0.01}$&0.84$^{+0.01}_{-0.01}$&0.82$^{+0.01}_{-0.01}$&0.82$^{+0.01}_{-0.01}$&0.88$^{+0.01}_{-0.02}$ \\
&$C_{\rm LE2}$&0.79$^{+0.01}_{-0.01}$&0.80$^{+0.01}_{-0.01}$&0.87$^{+0.01}_{-0.01}$&0.79$^{+0.02}_{-0.01}$&0.76$^{+0.02}_{-0.02}$ \\
&$C_{\rm ME0}$ &1 (fixed)&1 (fixed)&1 (fixed)&1 (fixed)&1 (fixed) \\
&$C_{\rm ME1}$&0.97$^{+0.01}_{-0.02}$&0.97$^{+0.01}_{-0.01}$&1.00$^{+0.02}_{-0.03}$&0.97$^{+0.01}_{-0.02}$&1.01$^{+0.03}_{-0.02}$ \\
&$C_{\rm ME2}$&1.03$^{+0.01}_{-0.01}$&1.02$^{+0.01}_{-0.01}$&0.89$^{+0.01}_{-0.01}$&1.02$^{+0.02}_{-0.01}$&0.99$^{+0.01}_{-0.02}$ \\
&$C_{\rm HE}$&1.20$^{+\rm p}_{-0.06}$&1.20$^{+\rm p}_{-0.04}$&1.16$^{+0.02}_{-0.04}$&1.20$^{+\rm p}_{-0.04}$&1.06$^{+0.03}_{-0.03}$ \\
\hline
GX 349+2&$\Gamma$&\nodata&2.16$^{+0.05}_{-0.04}$&\nodata&2.16$^{+0.17}_{-0.07}$&\nodata \\
&$N_{\rm H}$ (10$^{22}$ cm$^{-2}$)&0.53$^{+0.09}_{-0.06}$&0.92$^{+0.04}_{-0.04}$&0.72$^{+0.08}_{-0.08}$&0.92$^{+0.09}_{-0.06}$&0.62$^{+0.02}_{-0.02}$ \\
&$E_{\rm line}$ (keV)&6.96$^{+0.09}_{-0.06}$&6.97$^{+0.04}_{-0.04}$&7.03$^{+0.06}_{-0.12}$&7.03$^{+0.07}_{-0.10}$&6.76$^{+0.07}_{-0.08}$ \\
&$\sigma_{\rm line}$ (eV)&0.93$^{+0.06}_{-0.09}$&0.89$^{+0.06}_{-0.10}$&0.93$^{+0.07}_{-0.15}$&0.92$^{+0.05}_{-0.10}$&0.67$^{+0.14}_{-0.08}$ \\
&$T_{\rm in}$ (keV)&2.05$^{+0.02}_{-0.02}$&2.17$^{+0.02}_{-0.03}$&1.19$^{+0.03}_{-0.02}$&2.16$^{+0.02}_{-0.03}$&2.09$^{+0.02}_{-0.02}$ \\
&$kT_{\rm BB}$ (keV)&3.81$^{+0.16}_{-0.24}$&\nodata&1.75$^{+0.03}_{-0.03}$&\nodata&4.65$^{+0.21}_{-0.29}$ \\
\hline
&$\chi^2$ ($d.o.f.$)&2783.32 (2762)&2777.37 (2762)&2824.12 (2763)&2806.81 (2763)&2890.10 (2764)\\
\enddata
\tablecomments{$^*$: inner-calibration constants that significantly deviate from unit are caused by the uncertainty of the background model for faint sources. Based on joint observations with {\it NuSTAR} or other satellites, this will not dramatically influence the spectral shape \citep[for details, see][]{Chen2023}. We set an upper limit of 1.2 for these constant parameters to avoid for a large deviation. In the table, an upper error of ‘‘p’’ means that this parameter has reached its upper limit.
 \\Model 1: {\tt\string const*TBabs*TBpcf(gauss + bbodyrad + powerlaw*highecut)} for S$_1$ and {\tt\string const*TBabs(gauss + diskbb + bbodyrad)} for S$_2$.\\Model 2: {\tt\string const*TBabs*TBpcf(gauss + bbodyrad + powerlaw*highecut)} for S$_1$ and {\tt\string const*TBabs(gauss + diskbb + powerlaw)} for S$_2$.\\Model 3: {\tt\string const*TBabs*TBpcf(gauss + bbodyrad + cutoffpl)} for S$_1$ and {\tt\string const*TBabs(gauss + diskbb + bbodyrad)} for S$_2$.\\Model 4:  {\tt\string const*TBabs*TBpcf(gauss + bbodyrad + cutoffpl)} for S$_1$ and {\tt\string const*TBabs(gauss + diskbb + powerlaw)} for S$_2$. \\Model 5:  {\tt\string const*TBabs*TBpcf(gauss + bbodyrad + nthcomp)} for S$_1$ and {\tt\string const*TBabs(gauss + diskbb + bbodyrad)} for S$_2$. \label{model}}
\end{deluxetable}
\begin{figure}
	\includegraphics[width=0.5\linewidth]{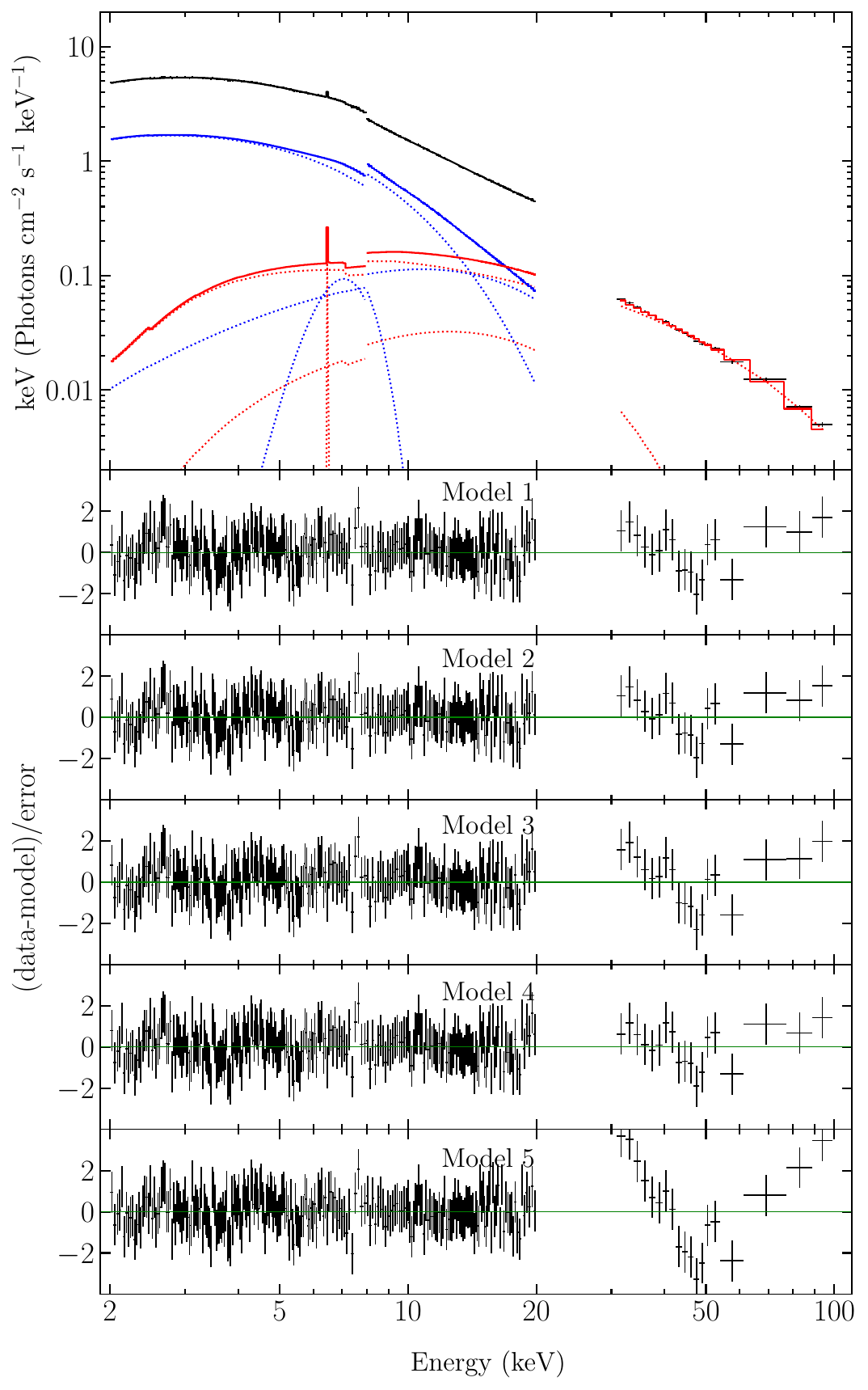}
	\centering
	\caption{Broadband spectral fitting of 4U 1700-37 and GX 349+2 using Model 1, where the red and blue lines indicate contributions from different sources and the black line is their summation. Each dotted line represents a component (e.g. {\tt \string cutoffpl}, {\tt \string bbodrad}) and the solid line gives the summation of each source.
	Lower panels show the residuals when using Models 1-5 (see Table~\ref{model}). 
	\label{Spectrum}}
\end{figure}
\begin{deluxetable}{ccccccc}
  \tablecaption{Best-fitting parameters of 4U 1700-37 and GX 349+2 with $\sim$ 50$\,$keV CRSF.  \label{para1}}
  \tabletypesize{\scriptsize}
  \tablewidth{400pt}
  \renewcommand\arraystretch{1.4}
  \tablehead{
  \colhead{Source} & \colhead{Parameter} & \colhead{Model 1} & 
  \colhead{Model 2} & \colhead{Model 3} & 
  \colhead{Model 4} &
  \colhead{Model 5} 
  } 
  \startdata
    4U 1700-37&$\Gamma$&1.06$^{+0.03}_{-0.03}$&1.16$^{+0.04}_{-0.07}$&0.85$^{+0.03}_{-0.03}$&0.91$^{+0.04}_{-0.03}$& 1.82$^{+0.03}_{-0.02}$ \\
    &${N_{\rm H,1}}$ (10$^{22}$ cm$^{-2}$)&0.50 (fixed)&0.50 (fixed)&0.50 (fixed)&0.50 (fixed)&0.50 (fixed) \\
    &$N_{\rm H,2}$ (10$^{22}$ cm$^{-2}$)&6.39$^{+0.31}_{-0.50}$&6.84$^{+0.72}_{-0.75}$&5.47$^{+0.49}_{-0.25}$&6.06$^{+0.44}_{-0.55}$& 3.16$^{+0.89}_{-0.69}$\\
    &$f$&0.98$^{+0.01}_{-0.01}$&0.95$^{+0.04}_{-0.07}$&0.99$^{+0.01}_{-0.08}$&0.96$^{+0.03}_{-0.06}$&1.00$^{+0.00}_{-0.14}$ \\
    &$E_{\rm line}$ (keV)&6.48$^{+0.02}_{-0.01}$&6.47$^{+0.04}_{-0.02}$&6.48$^{+0.02}_{-0.04}$&6.48$^{+0.02}_{-0.03}$&6.47$^{+0.03}_{-0.02}$ \\
    &${\sigma_{\rm line}}$ (eV)&10 (fixed)&10 (fixed)&10 (fixed)&10 (fixed)&10 (fixed) \\
    &$kT_{\rm BB}$ (keV)&3.04$^{+0.35}_{-0.21}$&3.74$^{+0.20}_{-0.28}$&2.91$^{+0.15}_{-0.08}$&3.08$^{+0.14}_{-0.10}$&\nodata \\
    &$Norm_{\rm BB}$ ($10^{-3}$)&405$^{+150}_{-79}$&319$^{+49}_{-68}$&1019$^{+95}_{-101}$&814$^{+73}_{-81}$&\nodata \\
    &$E_{\rm cut}$ (keV)&9.39$^{+0.38}_{-0.58}$&9.05$^{+0.77}_{-0.45}$&21.99$^{+0.72}_{-0.53}$&22.38$^{+2.02}_{-0.55}$&\nodata \\
    &$E_{\rm fold}$ (keV)&23.99$^{+1.26}_{-0.78}$&25.76$^{+1.37}_{-0.94}$&\nodata&\nodata&\nodata \\
    &$kT_{\rm e}$ (keV) &\nodata &\nodata &\nodata &\nodata &14.35$^{+0.44}_{-0.36}$ \\
    &$kT_{\rm seed}$ (keV) &\nodata &\nodata &\nodata &\nodata &1.92$^{+0.09}_{-0.03}$ \\
    &$Norm_{\rm comp}$ ($10^{-3}$) &\nodata &\nodata &\nodata &\nodata &19.11$^{+0.55}_{-1.21}$ \\
    &$Flux$ (10$^{-9}$\,erg\,cm$^{-2}$ s$^{-1}$)&6.77$^{+0.07}_{-0.12}$&6.69$^{+0.18}_{-0.10}$&6.78$^{+0.14}_{-0.12}$&6.67$^{+0.23}_{-0.03}$&6.77$^{+0.17}_{-0.15}$ \\
    &$E_{\rm cyc}$ (keV)&$49.56^{+1.38}_{-1.21}$&$49.40^{+11.29}_{-4.53}$&$50.26^{+1.69}_{-1.70}$&$49.33^{+5.15}_{-1.73}$&$50.62^{+2.72}_{-1.59}$ \\
    &$\sigma_{\rm cyc}$ (keV)&$10.19^{+3.06}_{-3.41}$&$8.88^{+9.49}_{-3.87}$&$10.41^{+3.78}_{-1.86}$&$10.39^{+3.44}_{-2.82}$&$10.24^{+2.72}_{-3.13}$ \\
    &$d_{\rm cyc}$&$4.45^{+1.69}_{-2.38}$&$2.84^{+6.44}_{-1.89}$&$5.22^{+3.43}_{-1.90}$&$4.90^{+2.24}_{-2.48}$&6.09$^{+2.59}_{-2.21}$ \\
    &Significance (AIC)&$4.1\,\sigma$&$2.7\,\sigma$&$6.9\,\sigma$&$5.4\,\sigma$&$8.9\,\sigma$ \\
    &Significance (simulation)&$>4\,\sigma$&$3.6\,\sigma$&$>4\,\sigma$&$>4\,\sigma$&$>4\,\sigma$ \\
    &$C_{\rm LE0}^*$ &0.78$^{+0.01}_{-0.01}$&0.79$^{+0.03}_{-0.02}$&0.77$^{+0.02}_{-0.01}$&0.78$^{+0.03}_{-0.01}$&0.77$^{+0.02}_{-0.02}$ \\
    &$C_{\rm LE1}$&0.85$^{+0.01}_{-0.01}$&0.85$^{+0.01}_{-0.01}$&0.84$^{+0.01}_{-0.01}$&0.84$^{+0.01}_{-0.01}$&0.87$^{+0.01}_{-0.01}$ \\
    &$C_{\rm LE2}$&0.77$^{+0.01}_{-0.01}$&0.78$^{+0.02}_{-0.02}$&0.74$^{+0.02}_{-0.01}$&0.77$^{+0.02}_{-0.01}$&0.76$^{+0.02}_{-0.02}$ \\
    &$C_{\rm ME0}$ &1 (fixed)&1 (fixed)&1 (fixed)&1 (fixed)&1 (fixed) \\
    &$C_{\rm ME1}$&0.99$^{+0.01}_{-0.02}$&0.99$^{+0.02}_{-0.03}$&1.00$^{+0.02}_{-0.03}$&1.00$^{+0.01}_{-0.04}$&1.02$^{+0.02}_{-0.03}$ \\
    &$C_{\rm ME2}$&1.01$^{+0.01}_{-0.01}$&1.01$^{+0.03}_{-0.02}$&1.00$^{+0.02}_{-0.02}$&1.00$^{+0.02}_{-0.01}$&0.99$^{+0.02}_{-0.02}$ \\
    &$C_{\rm HE}$&1.14$^{+0.04}_{-0.06}$&1.20$^{+\rm p}_{-0.10}$&1.13$^{+0.04}_{-0.05}$&1.19$^{+0.01}_{-0.09}$&1.08$^{+0.06}_{-0.05}$ \\
    \hline
    GX 349+2&$PhotonIndex$&\nodata&2.00$^{+0.26}_{-0.18}$&\nodata&1.94$^{+0.13}_{-0.05}$&\nodata \\
    &$N_{\rm H}$ (10$^{22}$ cm$^{-2}$)&0.55$^{+0.01}_{-0.01}$&0.84$^{+0.07}_{-0.08}$&0.52$^{+0.06}_{-0.09}$&0.82$^{+0.09}_{-0.06}$&0.61$^{+0.09}_{-0.12}$ \\
    &$E_{\rm line}$ (keV)&6.89$^{+0.05}_{-0.03}$&6.90$^{+0.12}_{-0.17}$&6.89$^{+0.05}_{-0.09}$&6.91$^{+0.09}_{-0.07}$&6.75$^{+0.09}_{-0.12}$ \\
    &$\sigma_{\rm line}$ (eV)&0.90$^{+0.06}_{-0.09}$&0.83$^{+0.12}_{-0.06}$&0.82$^{+0.13}_{-0.10}$&0.82$^{+0.09}_{-0.12}$&0.67$^{+0.19}_{-0.08}$ \\
    &$T_{\rm in}$ (keV)&2.04$^{+0.03}_{-0.02}$&2.16$^{+0.02}_{-0.02}$&2.08$^{+0.02}_{-0.02}$&2.15$^{+0.02}_{-0.02}$&2.03$^{+0.02}_{-0.02}$ \\
    &$kT_{\rm BB}$ (keV)&3.90$^{+0.26}_{-0.18}$&\nodata&4.19$^{+0.15}_{-0.23}$&\nodata&3.81$^{+0.13}_{-0.20}$ \\
    \hline
    &$\chi^2$ ($d.o.f.$)&2756.69 (2759)&2761.09 (2759)&2765.24 (2760)&2767.96 (2760)&2798.60 (2761)\\
  \enddata
  \tablecomments{Models 1-5 are the same as those in Table~\ref{model}.  \label{model2}}
\end{deluxetable}
\begin{deluxetable}{ccccccc}
  \tablecaption{Best-fitting parameters of 4U 1700-37 and GX 349+2 with $\sim$ 16$\,$keV CRSF.\label{para2}}
  \tabletypesize{\scriptsize}
  \tablewidth{400pt}
  \renewcommand\arraystretch{1.4}
  \tablehead{
  \colhead{Source} & \colhead{Parameter} & \colhead{Model 1} & 
  \colhead{Model 2} & \colhead{Model 3} & 
  \colhead{Model 4} &
  \colhead{Model 5} 
  } 
  \startdata
    4U 1700-37&$\Gamma$&1.29$^{+0.02}_{-0.05}$&1.32$^{+0.03}_{-0.03}$&1.16$^{+0.02}_{-0.03}$&1.21$^{+0.05}_{-0.03}$& 1.99$^{+0.06}_{-0.05}$ \\
    &${N_{\rm H,1}}$ (10$^{22}$ cm$^{-2}$)&0.50 (fixed)&0.50 (fixed)&0.50 (fixed)&0.50 (fixed)&0.50 (fixed) \\
    &$N_{\rm H,2}$ (10$^{22}$ cm$^{-2}$)&6.86$^{+0.65}_{-0.70}$&7.21$^{+0.54}_{-0.64}$&6.57$^{+0.17}_{-0.14}$&7.03$^{+0.44}_{-0.57}$& 3.19$^{+0.77}_{-0.53}$\\
    &$f$&0.92$^{+0.03}_{-0.03}$&0.94$^{+0.03}_{-0.04}$&0.96$^{+0.01}_{-0.01}$&0.93$^{+0.03}_{-0.04}$&1.00$^{+0.00}_{-0.17}$ \\
    &$E_{\rm line}$ (keV)&6.48$^{+0.03}_{-0.02}$&6.47$^{+0.02}_{-0.02}$&6.48$^{+0.01}_{-0.01}$&6.48$^{+0.02}_{-0.02}$&6.47$^{+0.04}_{-0.03}$ \\
    &${\sigma_{\rm line}}$ (eV)&10 (fixed)&10 (fixed)&10 (fixed)&10 (fixed)&10 (fixed) \\
    &$kT_{\rm BB}$ (keV)&4.31$^{+0.01}_{-0.12}$&4.39$^{+0.08}_{-0.05}$&4.13$^{+0.07}_{-0.05}$&4.16$^{+0.11}_{-0.11}$&\nodata \\
    &$Norm_{\rm BB}$ ($10^{-3}$)&504$^{+48}_{-67}$&429$^{+42}_{-47}$&716$^{+27}_{-48}$&635$^{+82}_{-56}$&\nodata \\
    &$E_{\rm cut}$ (keV)&8.99$^{+0.65}_{-1.03}$&8.66$^{+0.61}_{-1.56}$&28.74$^{+0.62}_{-0.68}$&29.28$^{+2.14}_{-1.62}$&\nodata \\
    &$E_{\rm fold}$ (keV)&30.63$^{+1.13}_{-1.62}$&31.17$^{+0.72}_{-0.77}$&\nodata&\nodata&\nodata \\
    &$kT_{\rm e}$ (keV) &\nodata &\nodata &\nodata &\nodata &16.44$^{+1.13}_{-0.90}$ \\
    &$kT_{\rm seed}$ (keV) &\nodata &\nodata &\nodata &\nodata &2.20$^{+0.09}_{-0.07}$ \\
    &$Norm_{\rm comp}$ ($10^{-3}$) &\nodata &\nodata &\nodata &\nodata &15.47$^{+0.75}_{-1.11}$ \\
    &$Flux$ (10$^{-9}$\,erg\,cm$^{-2}$ s$^{-1}$)&6.68$^{+0.12}_{-0.08}$&6.75$^{+0.15}_{-0.07}$&7.72$^{+0.04}_{-0.03}$&6.81$^{+0.05}_{-0.03}$&6.64$^{+0.09}_{-0.15}$ \\
    &$E_{\rm cyc}$ (keV)&$16.71^{+1.13}_{-0.74}$&$17.50^{+0.79}_{-0.52}$&$16.36^{+0.44}_{-0.16}$&$16.70^{+0.45}_{-0.38}$&16 (fixed) \\
    &$\sigma_{\rm cyc}$ (keV)&5 (fixed)&5 (fixed)&5 (fixed)&5 (fixed)&5 (fixed) \\
    &$d_{\rm cyc}$&$1.63^{+0.41}_{-0.35}$&$1.23^{+0.14}_{-0.16}$&$2.16^{+0.07}_{-0.10}$&$1.71^{+0.38}_{-0.47}$&2.24 $\times 10^{-7}$$^{+0.36}_{-\rm p}$ \\
    &Significance(\rm AIC) &$4.7\,\sigma$&$2.8\,\sigma$&$7.3\,\sigma$&$5.7\,\sigma$&$0.9\,\sigma$ \\
    &Significance (simulation)&$>4\,\sigma$&$>4\,\sigma$&$>4\,\sigma$&$>4\,\sigma$&$0.8\,\sigma$ \\
    &$C_{\rm LE0}^*$ &0.78$^{+0.02}_{-0.02}$&0.79$^{+0.02}_{-0.02}$&0.78$^{+0.01}_{-0.01}$&0.80$^{+0.02}_{-0.01}$&0.76$^{+0.01}_{-0.01}$ \\
    &$C_{\rm LE1}$&0.85$^{+0.01}_{-0.01}$&0.85$^{+0.01}_{-0.01}$&0.85$^{+0.01}_{-0.01}$&0.84$^{+0.01}_{-0.01}$&0.88$^{+0.01}_{-0.01}$ \\
    &$C_{\rm LE2}$&0.77$^{+0.02}_{-0.02}$&0.77$^{+0.01}_{-0.01}$&0.88$^{+0.01}_{-0.02}$&0.79$^{+0.01}_{-0.01}$&0.76$^{+0.02}_{-0.01}$ \\
    &$C_{\rm ME0}$ &1 (fixed)&1 (fixed)&1 (fixed)&1 (fixed)&1 (fixed) \\
    &$C_{\rm ME1}$&1.00$^{+0.02}_{-0.03}$&0.99$^{+0.02}_{-0.02}$&0.99$^{+0.01}_{-0.01}$&0.98$^{+0.01}_{-0.01}$&1.01$^{+0.02}_{-0.02}$ \\
    &$C_{\rm ME2}$&1.00$^{+0.02}_{-0.01}$&1.01$^{+0.02}_{-0.02}$&1.01$^{+0.01}_{-0.01}$&1.02$^{+0.01}_{-0.01}$&0.99$^{+0.01}_{-0.01}$ \\
    &$C_{\rm HE}$&1.20$^{+\rm p}_{-0.05}$&1.20$^{+\rm p}_{-0.06}$&1.20$^{+ \rm p}_{-0.05}$&1.20$^{+ \rm p}_{-0.06}$&1.07$^{+0.05}_{-0.05}$ \\
    \hline
    GX 349+2&$PhotonIndex$&\nodata&2.02$^{+0.09}_{-0.09}$&\nodata&2.06$^{+0.08}_{-0.07}$&\nodata \\
    &$N_{\rm H}$ (10$^{22}$ cm$^{-2}$)&0.55$^{+0.03}_{-0.02}$&0.85$^{+0.06}_{-0.06}$&0.56$^{+0.01}_{-0.01}$&0.87$^{+0.06}_{-0.08}$&0.63$^{+0.03}_{-0.02}$ \\
    &$E_{\rm line}$ (keV)&6.86$^{+0.10}_{-0.09}$&6.91$^{+0.12}_{-0.09}$&6.91$^{+0.05}_{-0.05}$&6.95$^{+0.05}_{-0.09}$&6.76$^{+0.06}_{-0.08}$ \\
    &$\sigma_{\rm line}$ (eV)&0.93$^{+0.06}_{-0.11}$&0.86$^{+0.11}_{-0.10}$&0.96$^{+0.04}_{-0.07}$&0.90$^{+0.07}_{-0.10}$&0.68$^{+0.12}_{-0.10}$ \\
    &$T_{\rm in}$ (keV)&2.03$^{+0.03}_{-0.03}$&2.16$^{+0.02}_{-0.02}$&2.03$^{+0.02}_{-0.01}$&2.15$^{+0.01}_{-0.01}$&2.09$^{+0.02}_{-0.02}$ \\
    &$kT_{\rm BB}$ (keV)&3.86$^{+0.20}_{-0.20}$&\nodata&3.81$^{+0.11}_{-0.08}$&\nodata&4.70$^{+0.26}_{-0.12}$ \\
    \hline
    &$\chi^2$ ($d.o.f.$)&2753.89 (2760)&2760.67 (2760)&2757.89 (2761)&2766.06 (2761)&2890.08 (2763)\\
  \enddata
  \tablecomments{Models 1-5 are the same as those in Table \ref{model}.
  \label{model3}}
\end{deluxetable}
\begin{figure}
	\includegraphics[width=0.9\linewidth]{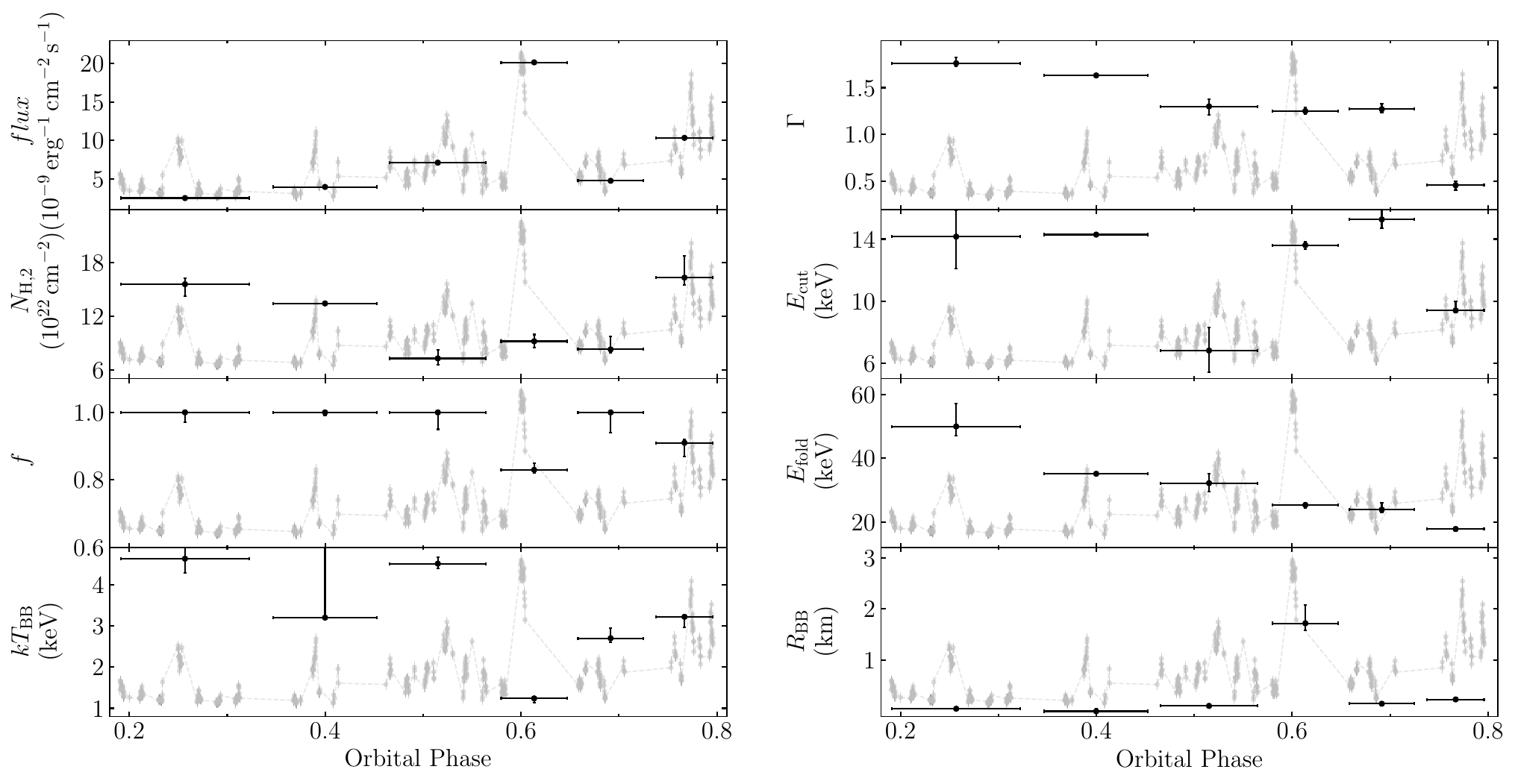}
	\centering
	\caption{
	Spectral evolution during the out-of-eclipse state of 4U 1700-37 using the Model 1 listed in Table~\ref{model}. 
    The grey points are HE lightcurves shown in Figure~\ref{lightcurve}.
		\label{Spectral_parameters}}
\end{figure}
\section{discussion} \label{sec4}
We performed a timing and spectral analysis of the HMXB 4U 1700-37 observed with $Insight$-HXMT during the out-of-eclipse orbital phase 0.19-0.79.
We find that 4U 1700-37 shows a high variability in flux on the time scale of kilo-seconds, similar to other SgXBs, e.g., 4U 1907+09, Vela X-1 and GX 301-2 \citep{Doroshenko(2012), Odaka(2013), Islam(2014)}.
This might be just due to variations in accretion efficiency, as stellar winds from the companion could be in-homogeneous and/or clumpy \citep[e.g.,][]{Ducci2009}.
%
Alternatively, star wind material ionized by X-rays may form a temporary accretion disk because of the Kelvin-Helmholtz instability in the wake region \citep{Fryxell(1988)}, leading to enhanced accretion rate, i.e., the flares that we observed.
Using a {\it Suzaku} observation, \citet{Jaisawal(2015)} reported a low flux segment and an increase in the column density in the 0.63–0.73 orbital phase, which was interpreted as the presence of accretion wake.
This decrease in flux was also seen in the averaged orbital intensity profile observed with {\it RXTE}-All Sky Monitor \citep{Islam2016}.
In our observations, the trends of flux and absorption are similar to the results of {\it Suzaku}, which also supports the accretion wake scenario.

%
%
%
%
%
%

The absorption $N_{\rm H,2}$ is relatively stable throughout the orbital phase, independent of the flux, which can be naturally understood if the accretion is via an accretion disk.
The presence of the accretion disk is potentially supported by the discovery of QPOs.
However, in our observation, we did not find significant QPOs, although the 7\,mHz signal could be a hint.
In literature, QPOs in 4U 1700-37 have been observed occasionally within the frequency range of 0.63-20 mHz \citep[e.g.,][]{Boroson(2003),Dolan(2011),Jaisawal(2015),Martinez-Chicharro(2018)}.
However, we caution that all these detections have a low significance level (i.e., $\lessapprox$3$\sigma$), and thus need to be confirmed by further observations.
If the QPOs are real, they can be used to infer the nature of the compact object by using the correlation between the QPO frequency and the luminosity, as suggested by different theoretical models \citep[e.g.,][]{Tagger1999, Ingram2009, Alpar1985, Klis1987, Shirakawa2002, Abolmasov2020}.

We studied the broadband spectroscopy of 4U 1700-37.
In general, it can be well described as a phenomenological {\tt\string cutoffpl/highecut} model or a physical {\tt\string nthcomp} model.
In addition, we detected hints of CRSFs around $\sim$16\,keV and $\sim$50\,keV.
The former is in well agreement with the result of \citet{Bala(2020)}, and the latter is first reported.
We caution that although 50\,keV is approximately three times than 16\,keV, they cannot be regarded as the fundamental and harmonics, because these two lines could not be constrained simultaneously.

If we assume that both lines are real, the energy ratio of $\sim$ 3 is much larger than the expected 2 between the fundamental and the first harmonic.
This could be caused by a non-dipole magnetic field, the distortion of the local magnetic field or multiple line-forming regions \citep{Nishimura2005, Furst2018,Mukherjee2012}.

Furthermore, significance levels for both lines are marginal and model-dependent, and therefore they need to be confirmed in the future.
On the other hand, the 37\,keV CRSF, suggested by \citet{Reynolds(1999), Jaisawal(2015)}, was not discovered.
If it exists, the non-detection might be due to the fact that 1) the absorption feature, is very weak\footnote{For example, the statistical probability of including this line by chance is up to 3\% \citep{Jaisawal(2015)}.};
2) the contamination source and the gap between 20-30\,keV caused by the high background inevitably affect the sensitivity of detecting fine spectral structures.
Nevertheless, we note that spectral parameters of the continuum we found is well consistent with previous studies, which confirms that \textit{Insight}-HXMT is capable of conducting spectral analysis and obtaining reliable results under a stable contamination.
At high energies, we did not discover the hard tail commonly observed in black hole binaries \citep{McConnell2002,CadolleBel2006,Cangemi2021}.
In addition, no significant spectral evolution is found during the flaring, during which the spectrum is still dominated by a non-thermal component, unlike the case of black holes that we observed previously \citep[see, e.g.,][]{Remillard2006}.
This suggests that the compact object is more likely to be a neutron star rather than a black hole.
On the other hand, if the source is a weakly magnetized neutron star, as the case in low mass X-ray binaries, significant thermal radiation from the surface of the neutron star or/and the boundary layer that leads to the tracking in the color-color diagram would be expected when the source is bright \citep[e.g.,][]{Mitsuda1989, Popham2001}, which however was not observed.
Another possibility is that the source is an accreting pulsar. 
In this case, its spectral shape is indeed well consistent with other systems.
However, it is puzzling why there are no pulsations, unless the rotational and magnetic axes are perfectly aligned or the source has a prolonged pulse period beyond the frequency range we detected \citep[e.g., some cases in the catalogue provided by][]{Neumann(2023)}.
Therefore, we conclude that new methods and further observations, such as the radio-X-ray fundamental plane study \citep{Gallo2018,vandenEijnden2022} or the polarization study \citep{Krawczynski2022, Doroshenko2022}, are still needed in order to determine the nature of the source.
\\

This work is supported by the National Natural Science Foundation of China under grants Nos. 12173103, U2038101, U1938103, 11733009, and 12261141691. 
This work is based on observations with {\it Insight}-HXMT, a project funded by the China National Space Administration (CNSA) and the Chinese Academy of Sciences (CAS). 

\newpage
\appendix
\renewcommand\thefigure{\thesection.\arabic{figure}}    
\setcounter{figure}{0}  
\section{HE background}
In Figure~\ref{background}, we present an unfolded spectrum (background-subtracted) and its background estimated by using the tool {\sc hebackmap} in HXMTdas v2.05. 
\begin{figure}[h]
	\includegraphics[width=0.5\linewidth]{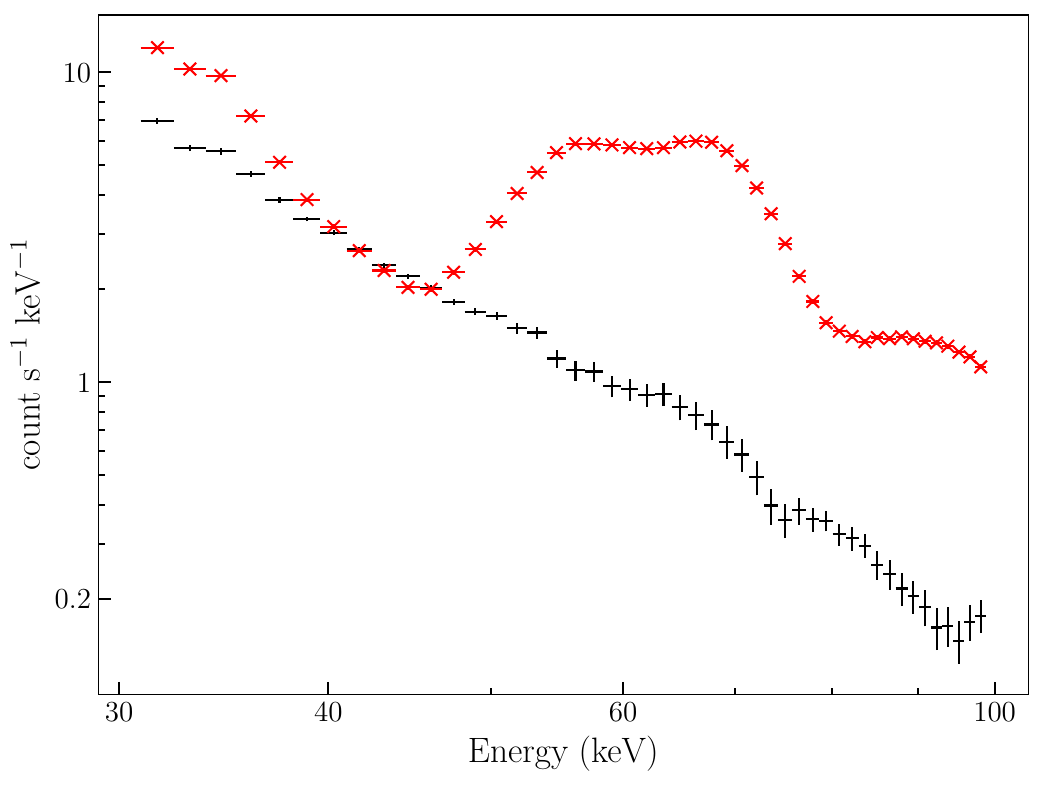}
	\centering
	\caption{A representative background-subtracted spectrum (black) and the corresponding background (red) observed with {\it Insight}-HXMT/HE.
	\label{background}}
\end{figure}

\newpage
\section{Residual with and witout CRSF}
In Figure \ref{fig:residual}, we show the residuals with different models after including the cyclotron lines and setting line depth to zero, respectively.
\begin{figure}[h]
  \centering
    \includegraphics[width=0.45\linewidth]{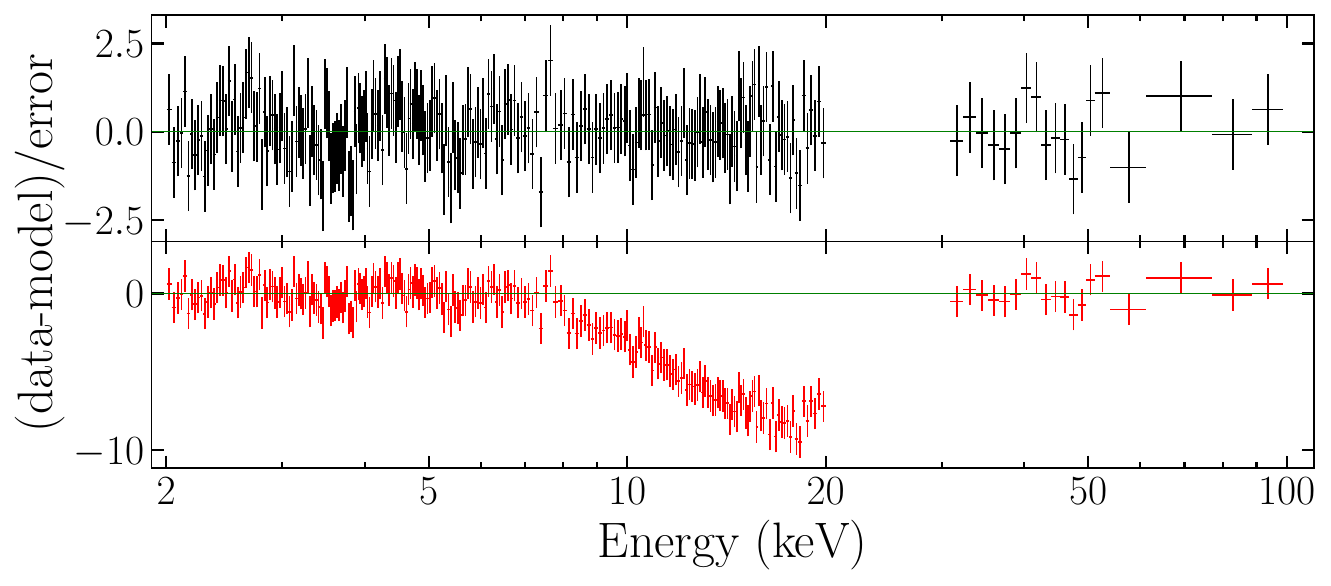}
    \includegraphics[width=0.45\linewidth]{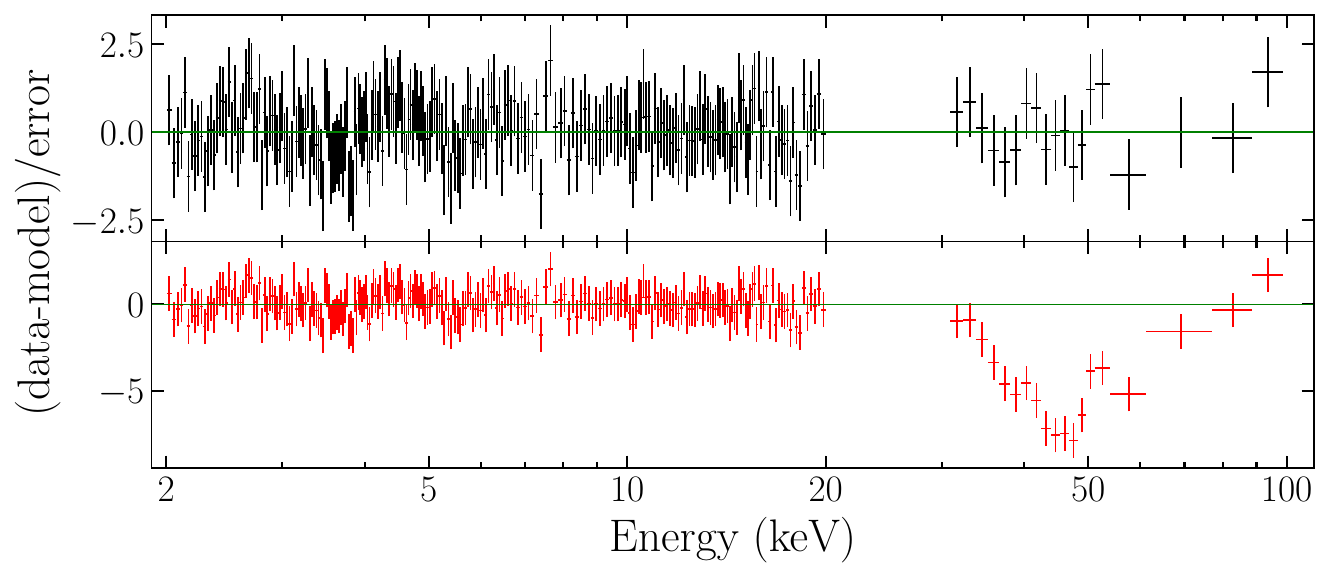}
    \includegraphics[width=0.45\linewidth]{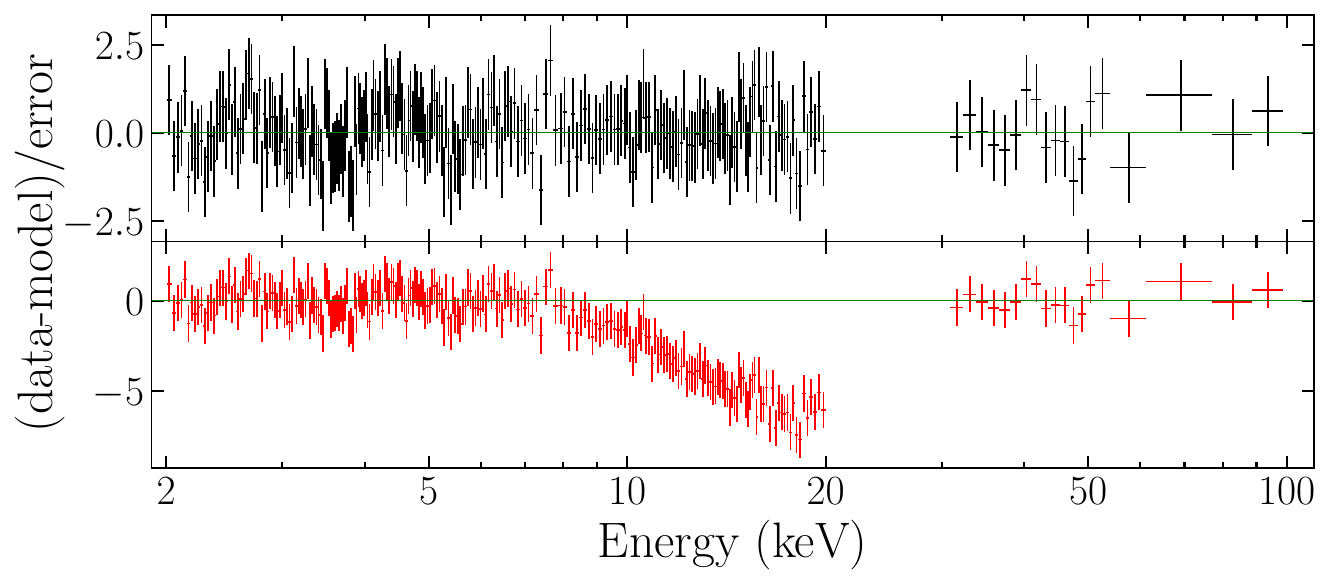}
    \includegraphics[width=0.45\linewidth]{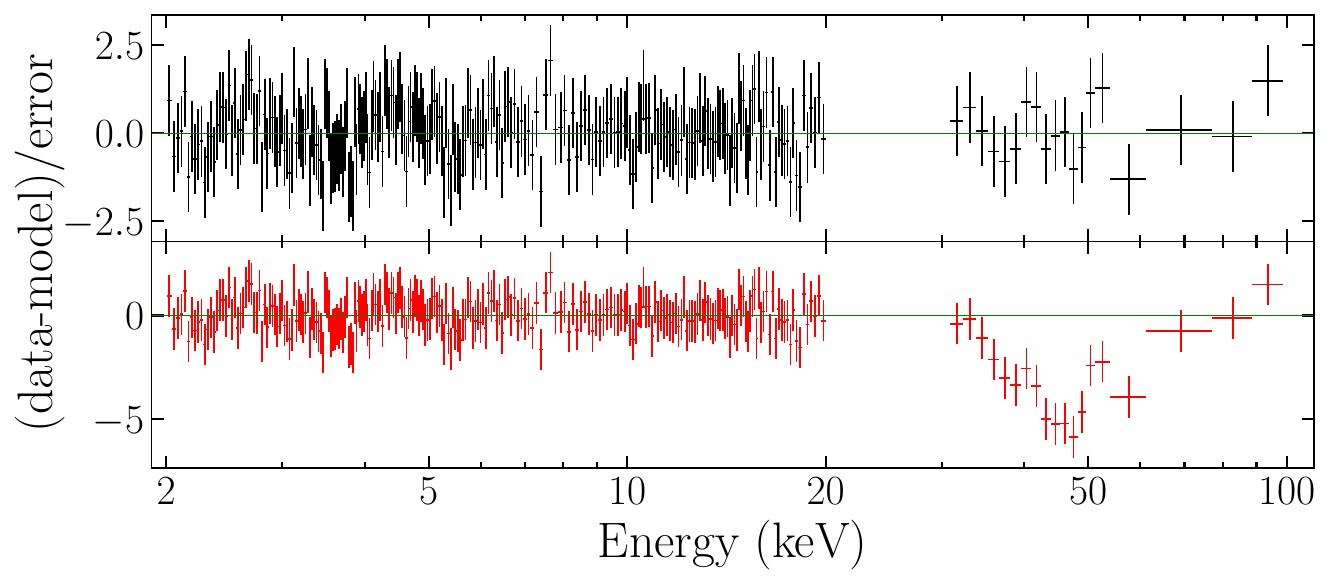}
    \includegraphics[width=0.45\linewidth]{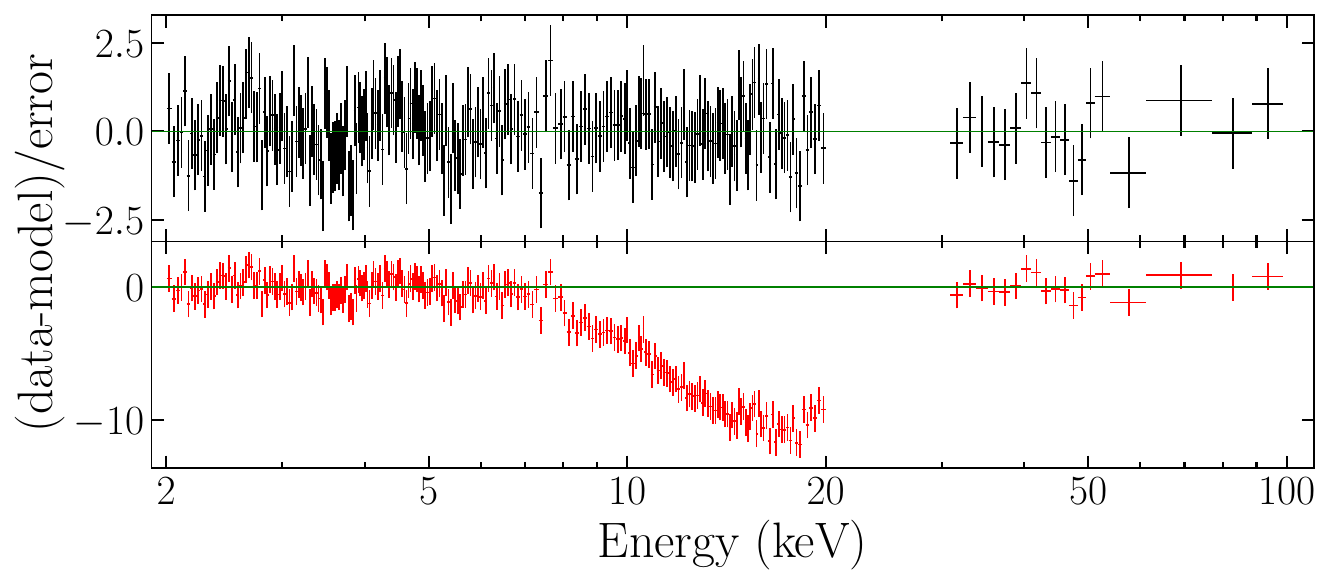}
    \includegraphics[width=0.45\linewidth]{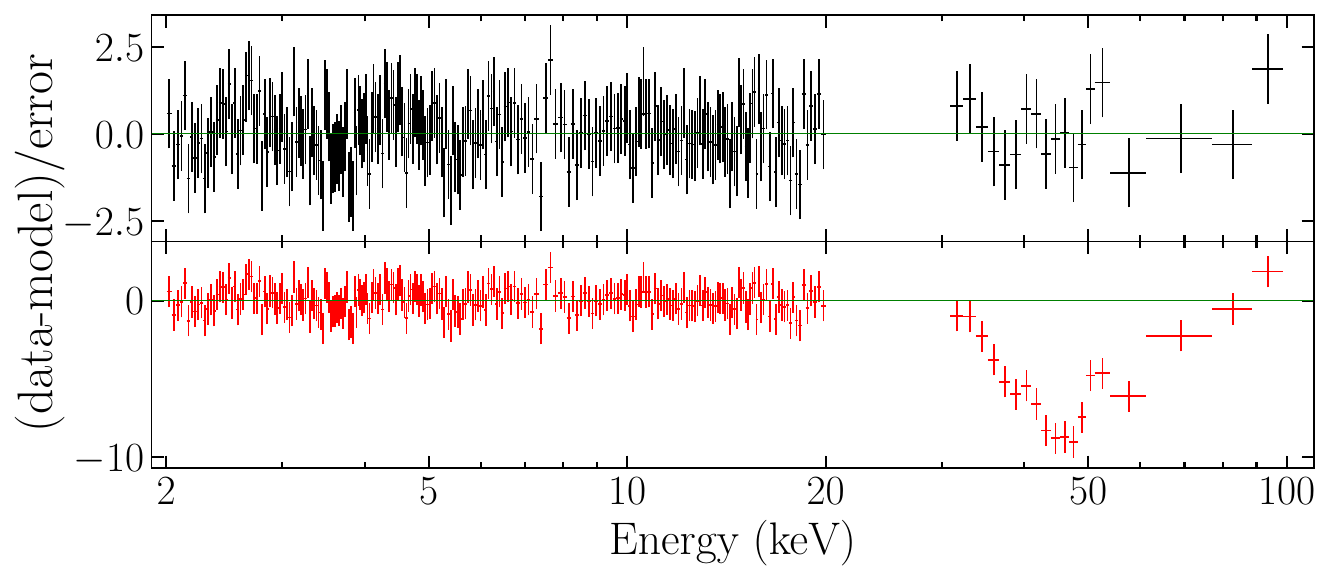}
    \includegraphics[width=0.45\linewidth]{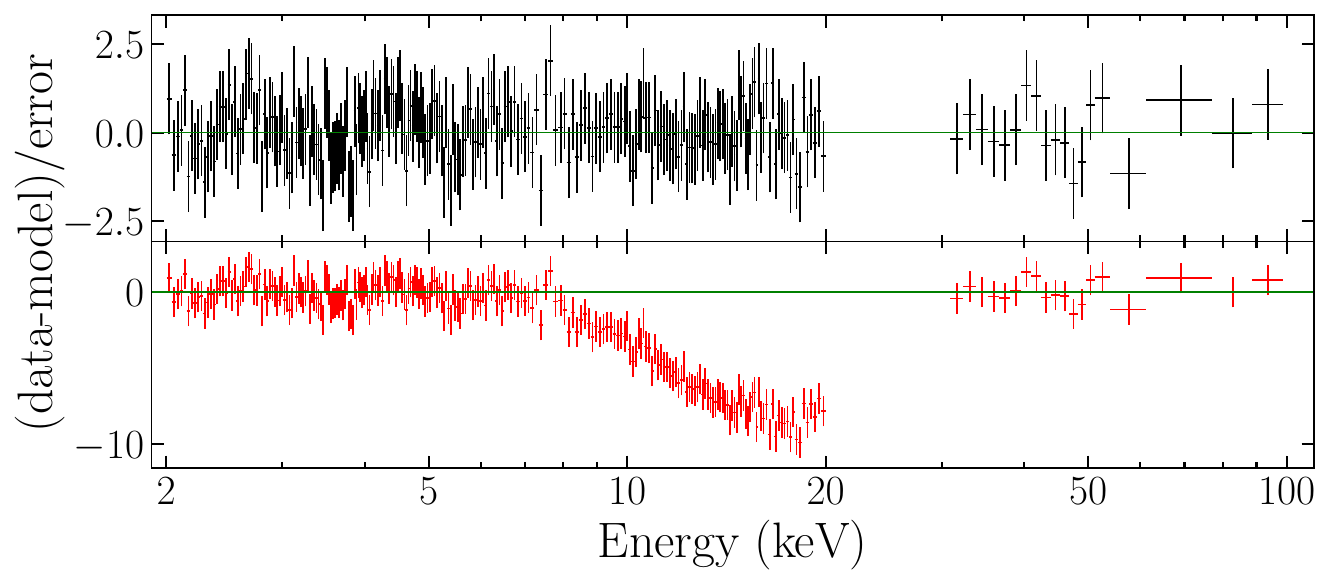}
    \includegraphics[width=0.45\linewidth]{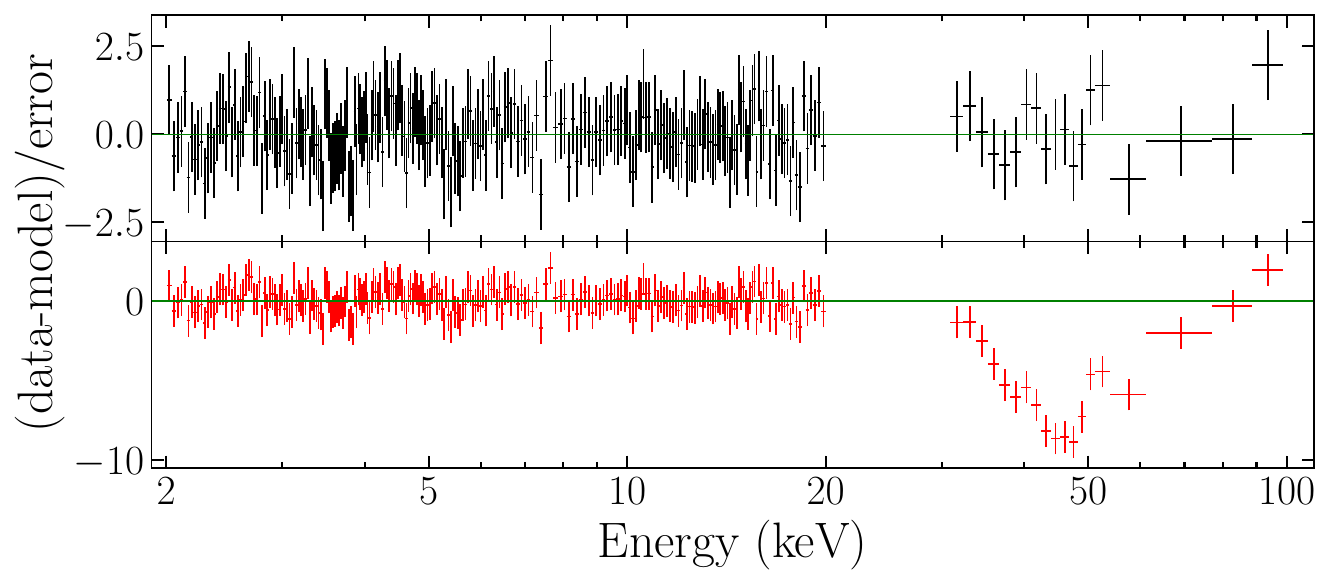}
    \includegraphics[width=0.45\linewidth]{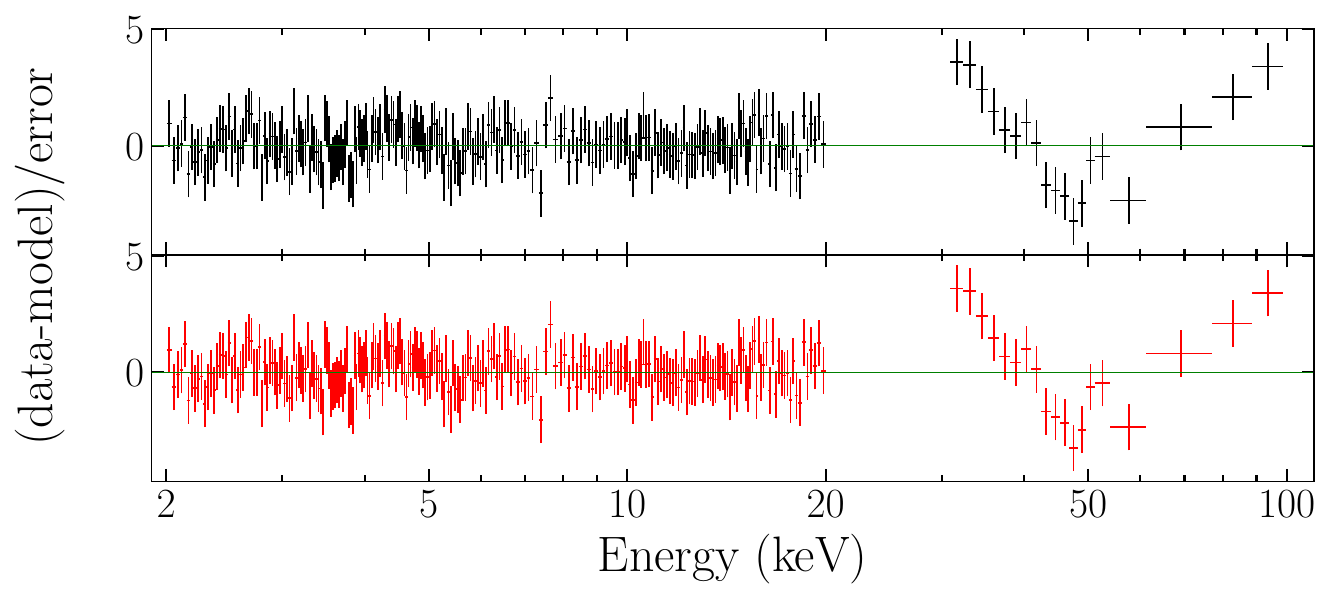}
    \includegraphics[width=0.45\linewidth]{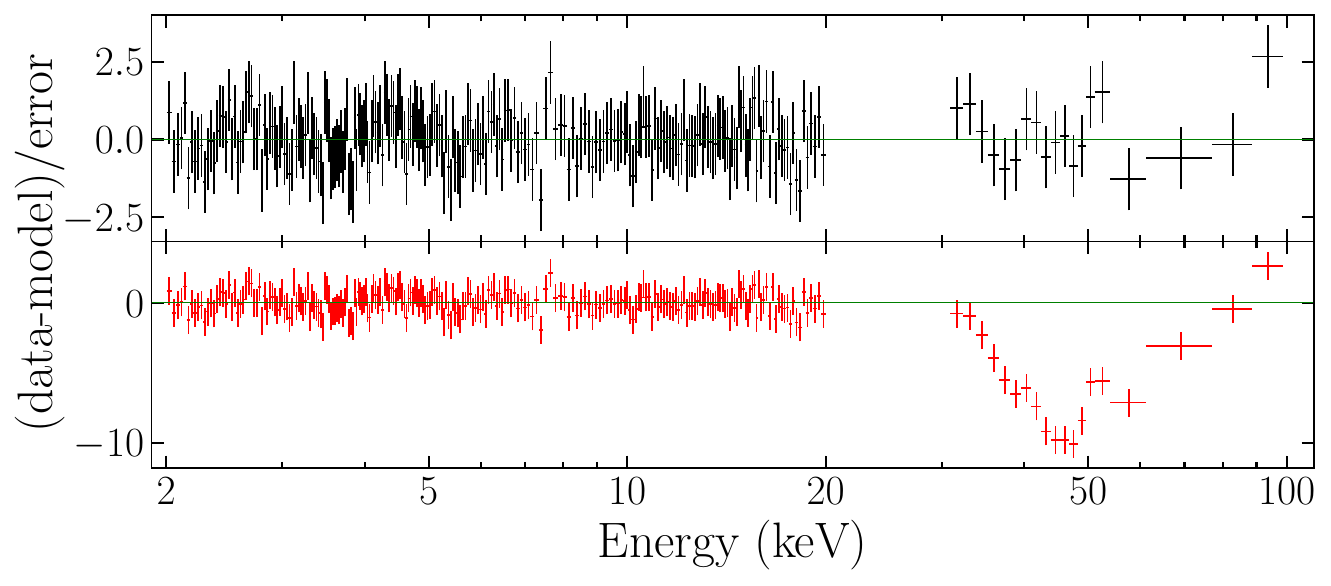}
  \caption{
  Black lines show best-fitting residuals of different continuum models (from top to bottom, model 1 to 5) with the 16\,keV (left) and 50\,keV (right) CRSF, and red lines present the corresponding residuals by removing CRSFs.
  }
  \label{fig:residual}
\end{figure}

\bibliography{sample631}{}
\bibliographystyle{aasjournal}
\end{document}